\documentclass[doublecol]{epl2} 

\usepackage{subfigure}
\usepackage{url}
\usepackage{mathpazo}
\usepackage{amsmath, amssymb, bm, textcomp}
\usepackage{booktabs} 
\usepackage[breaklinks]{hyperref}
\usepackage[a4paper,inner=2cm,outer=2cm,top=2.5cm,bottom=2cm]{geometry}

\usepackage{url}

\title{Age-structured estimation of COVID-19 ICU demand from low quality data}
\shorttitle{Estimation of COVID-19 ICU demand} 

\author{R. Veiga \inst{1} \and R. Murta\inst{2} \and R. Vicente\inst{3,4}}
\shortauthor{R. Veiga \etal}

\institute{                    
  \inst{1} Instituto de F\'{i}sica - Universidade de S\~{a}o
Paulo,  05508-090, S\~{a}o Paulo-SP, Brazil\\
  \inst{2} Looqbox - 04547-130, São Paulo-SP, Brazil\\
  \inst{3} Experian DataLab LatAm  - 04547-130, S\~{a}o Paulo-SP, Brazil \\
   \inst{4} Instituto de Matem\'{a}tica e Estat\'{i}stica, Universidade de S\~{a}o
Paulo -  05508-090, S\~{a}o Paulo-SP, Brazil
}


\abstract{We sample aggravated cases following age-structured probabilities from confirmed cases and use ICU occupation data to find a subnotification factor. A logistic fit is then employed to project the progression of the COVID-19 epidemic with plateau scenarios taken from locations that have reached this stage. Finally, the logistic curve found is corrected by the subnotification factor and sampled to project the future demand for ICU beds. }

\begin{document}

\maketitle

\section{Introduction}

The COVID-19 pandemic is ravaging the world and requiring every research energy available to help local public administrators dealing with the crisis. Brazil, unfortunately, is an emblematic case of a public health emergency mismanagement. Despite many voluntary initiatives  \cite{obs_covid_br, covid_radar, brasil_io,UFMG_labdec, Costa2020.05.06.20093492, yang2020modeling, Canabarro2020.04.03.20052498, RochaFilho2020.03.14.20035873, Alves2020.05.20.20108415}, the country lacks publicly available data that is complete, consistent and timely to monitor the pace of the epidemic.

 A main concern in many locations like Brazil is how to use incomplete data of low quality to anticipate the demand for crucial and limited Intensive Care Units (ICU). 

Compartment based epidemiological models \cite{diekmann, brauer2019mathematical}, like SIR, SEIR or many other more realistic variants, require the estimation of a number of parameters. Any future scenarios derived from the equations defining these models are critically dependent on these parameters that, in their turn, depend on the quality of the data available. Having data of admittedly low quality, makes the task of fitting realistic models questionable, at best.

The situation is further complicated by Sars-CoV-2 being a new virus with uncertain epidemiological parameters and by the complexities of severely unequal societies. To overtake these limitations, here we use the fact that the first wave of epidemics has already been resolved in several locations to inform the proposition of sensible and simple scenarios based only on very generic, yet robust, dynamical features. We then use clinical data from those same locations, data on ICU utilization and demographic data to estimate ICU demand by Monte Carlo simulation \cite{binder_book}.

\section{Building scenarios}

We start by considering the evolution in time of the number of confirmed cases. A very general feature, captured by compartment models, is that there is an initial exponential growth followed by a plateau, eventually reached when the number of susceptible declines. The simplest structure like that is provided by a logistic function. We thus model the evolution of confirmed cases $\hat{n}(t)$ for $t>0$ as \cite{Wu_2020}:
\begin{equation}
\label{eq:sig}
    \hat{n}(t) =  \frac{\hat{n}^{*}}{1 + e^{-\alpha(t - t_0)}}  \;,
\end{equation}
where $\alpha$ is the rate of the early exponential growth and $\hat{n}^{*}$ is the number of cases attained when the epidemic hits the plateau for $ \alpha (t - t_0)  \gg 1$. The time shift $t_0$ marks the inflection point $n(t_0) = \hat{n}^{*} /2 $.

\begin{figure*}[!htbp]
\subfigure[Confirmed COVID-19 cases and deaths in Brazil \cite{brasil_io}. \label{fig:BRA_COVID_}]{
    \includegraphics[scale=0.55]{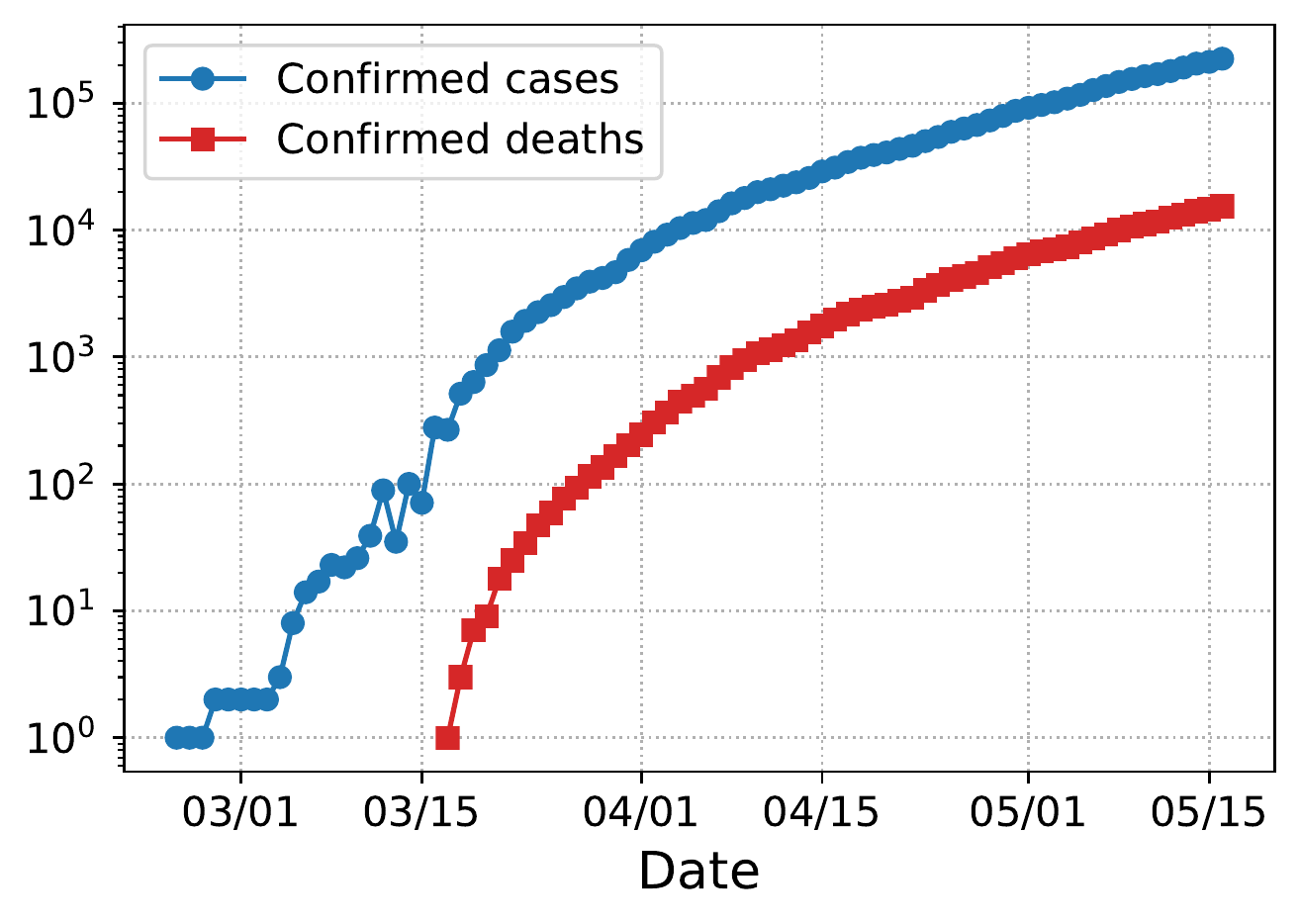}
    }
    \hspace{1cm}
\subfigure[Confirmed COVID-19 cases in S\~{a}o Paulo \cite{brasil_io}. \label{fig:SP_COVID}]{
    \includegraphics[scale=0.55]{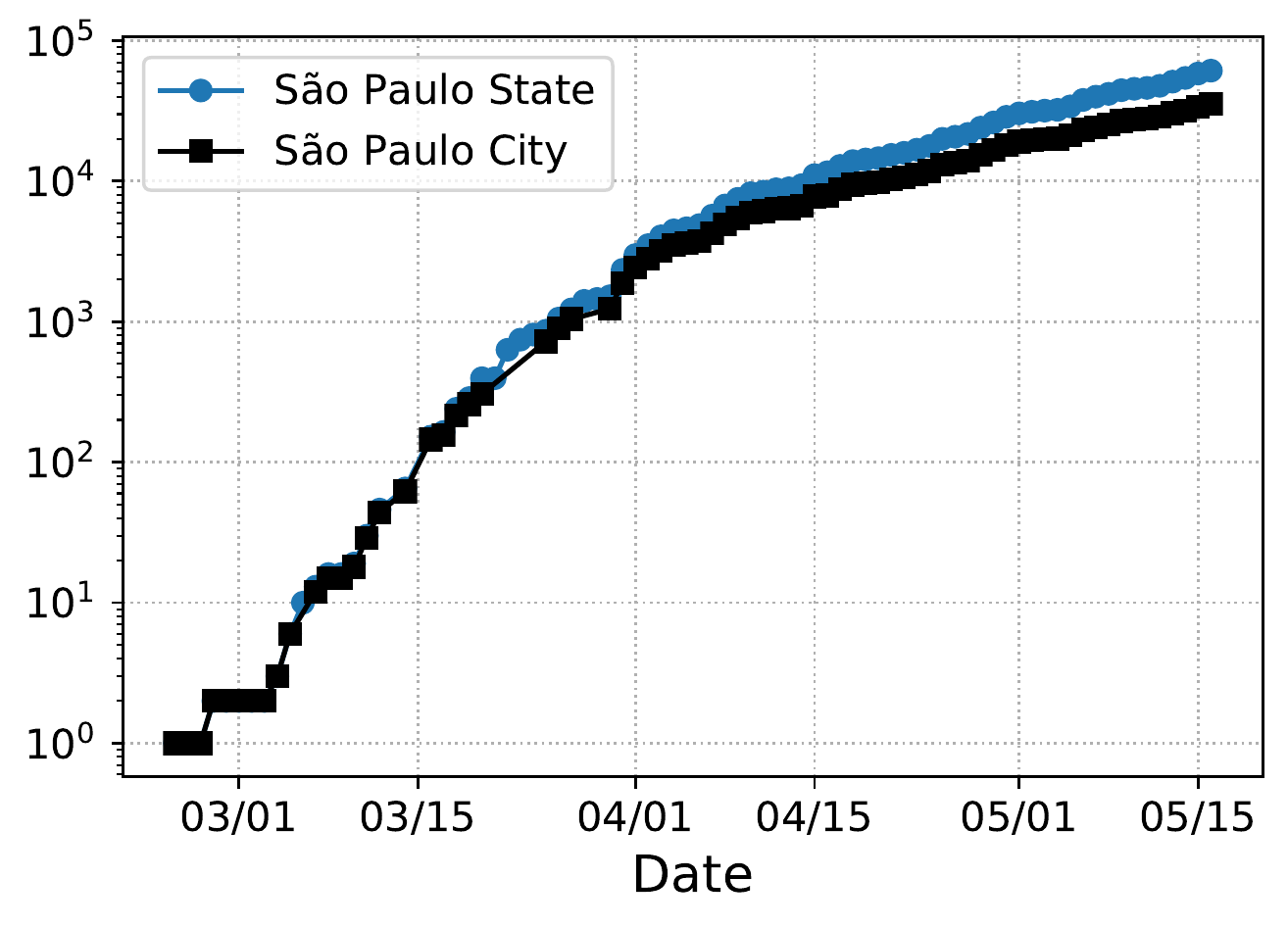}
    }
    \caption{Epidemic dynamics: confirmed cases.}
\end{figure*}

Figure \ref{fig:BRA_COVID_} represents Brazilian official records for confirmed cases and deaths as they were presented at May 17, 2020. It can be verified that, at this date, the epidemic hadn't plateaued yet. 

We focus our analysis at the state of S\~{a}o Paulo, the most populous state ($\approx$ 21\% of Brazilian population) and also the state housing the megalopolis that is the epidemic epicenter: the city of S\~{a}o Paulo ($\approx 27\%$ of S\~{a}o Paulo state's population).  Figure \ref{fig:SP_COVID} depicts confirmed cases for the state and the city of S\~{a}o Paulo until the same May 17, 2020. We do that because for S\~{a}o Paulo we have daily ICU occupation reports.

In order to find plausible values for $ \hat{n}^{*} $ we look at the epidemic dynamics across all countries affected and calculate relative contagion velocities $ \beta (t) / \hat{n}(t)$ , with 

\begin{equation}
\label{eq:EWMA}
\beta (t) = a  \Delta_{k} \hat{n} (t) + (1-a) \beta(t - 1)  \;,
\end{equation}
for k-lagged first differences
\begin{equation}
\label{eq:delta}
\Delta_k  \hat{n}(t)=  \frac{\hat{n}(t+k) - \hat{n}(k)}{k}  \;,
\end{equation}
where we have used $a=0.2$ and $k=5$. We then adopt to neglect countries where the relative contagion speed is greater than 0.5 \% as a criteria to identify locations that have reached an epidemic plateau. We further restrict our search to countries with total population from five to tens of million inhabitants \footnote{Code available at \url{https://github.com/rodsveiga/ICU_demand}}. For each country $L$ that have already reached the epidemic plateau we considered the number of cases per inhabitant as providing a different scenario, to say,
\begin{equation}
    \hat{n}^{*}_{L} = \frac{\hat{n}_{L} (t_{c})}{ N_L } N \;,
\end{equation}
where $t_c$  time when $ \beta  \le \beta_c$, $N$ and $N_L$ are, respectively, the populations we want to model and the population of the country that provides the scenario. 

With this simple approach we have selected Switzerland to provide an ``optimistic" scenario and Spain to provide a ``pessimistic" scenario. Table \ref{tab:scenarios} lists the number of cases expected in the plateau for the each scenario.


\begin{table}[h!]
\centering
\begin{tabular}{| c c c |}
\hline
 $L$ & $\hat{n}_{L} (t_{c})$ & $N_L$\\  
 \hline\hline
Spain & 229,047 & 46,795,540 \\
\hline
Switzerland & 163,071  &  8,513,227 \\
\hline
\end{tabular}
\caption{ \label{tab:scenarios} Scenarios for $\hat{n}^{*}$. Total population provided by \cite{worldbank}.} 
\end{table}



\begin{figure*}[!ht]
\subfigure[Expected epidemic development. State of S\~{a}o Paulo. Parameters: $\alpha_{\text{Spain}} = 0.0535$, $t_{0,\text{Spain}} = 95$ days;  $\alpha_{\text{Swiss}} = 0.0683$,  $t_{0,\text{Swiss}} = 87$ days. \label{fig:FITs_state_SP}]{
    \includegraphics[scale=0.55]{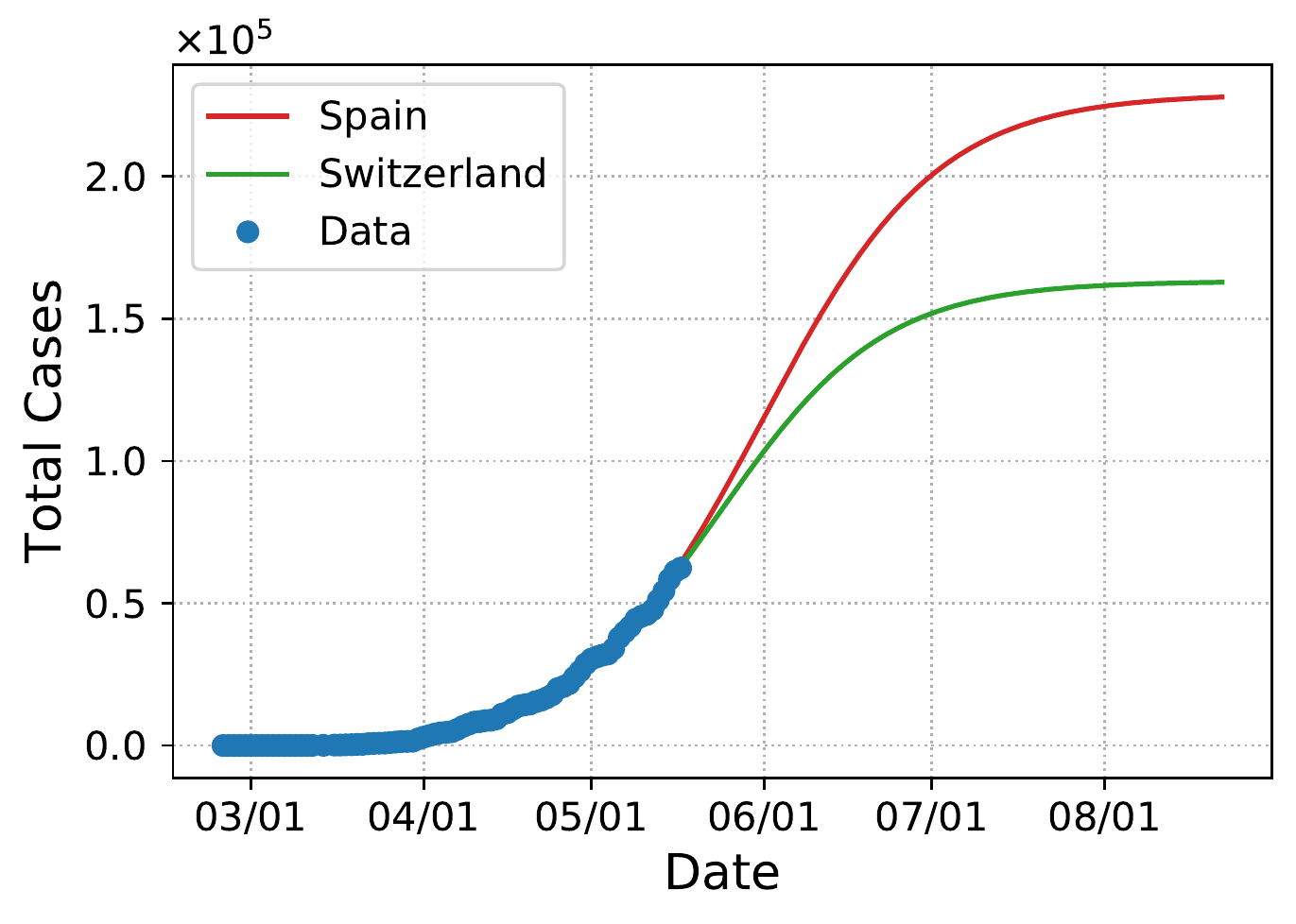}
    }
    \hspace{0.8cm}
\subfigure[Expected epidemic development. City of S\~{a}o Paulo. Parameters: $\alpha_{\text{Spain}} = 0.0742$,  $t_{0,\text{Spain}} = 68$ days; $\alpha_{\text{Swiss}} = 0.0986$,  $t_{0,\text{Swiss}} = 60$ days. \label{fig:FITs_city_SP}]{
    \includegraphics[scale=0.55]{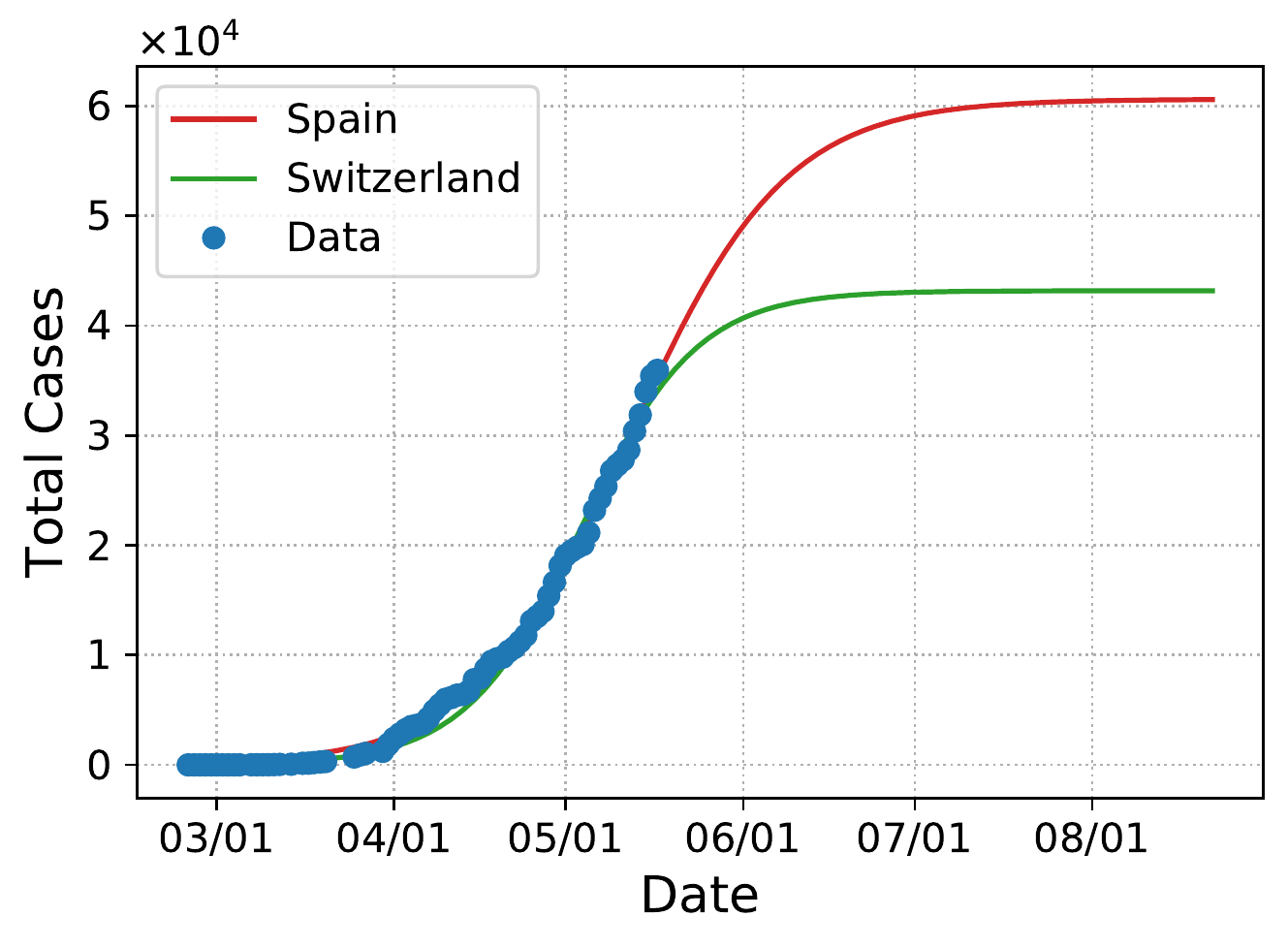}
    }
    
\subfigure[Age-structured population fraction in S\~{a}o Paulo State \cite{IBGE_age_SP}.   \label{fig:age_demography}]{
    \includegraphics[scale=0.55]{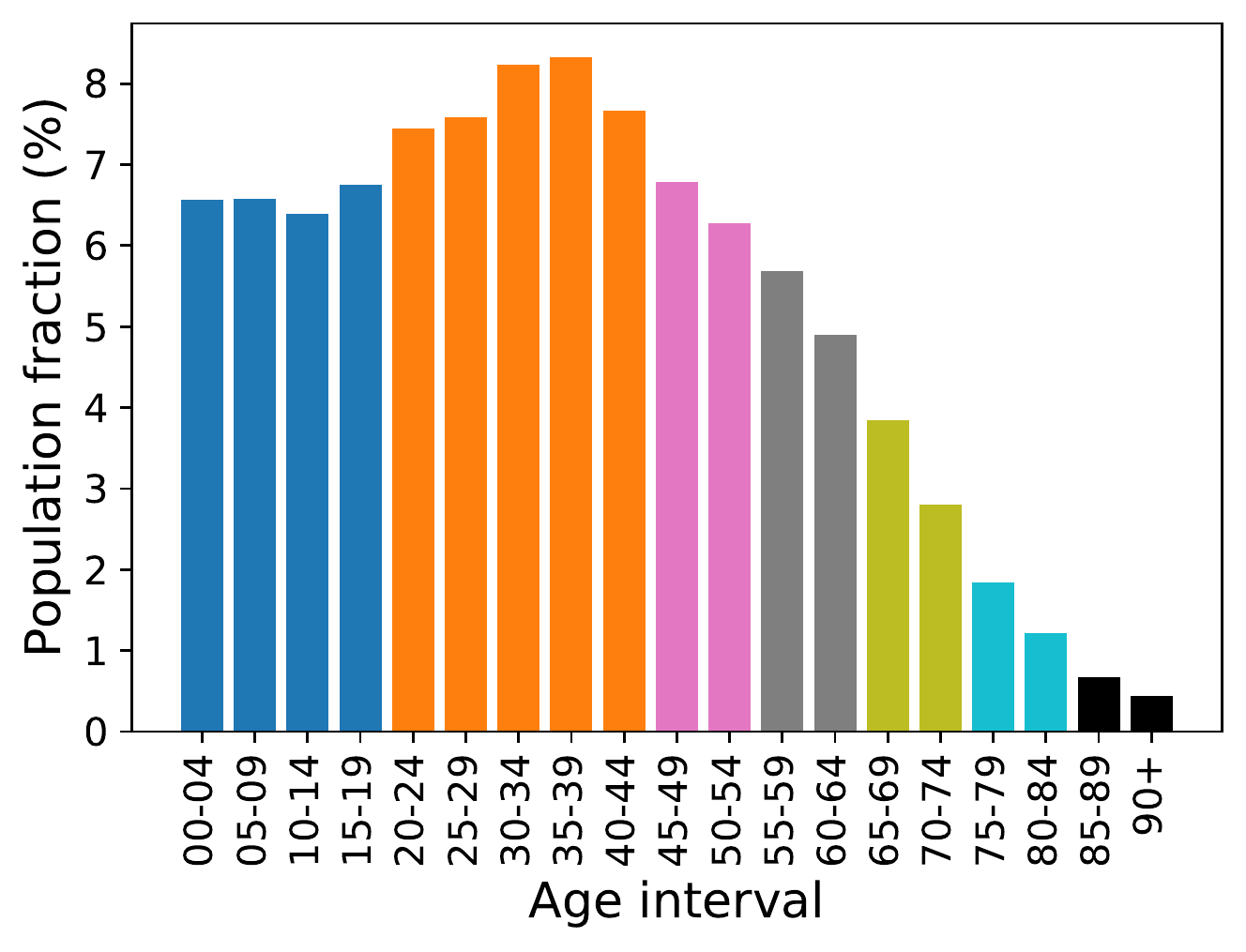}
    }
    \hspace{0.8cm}
\subfigure[ICU admission by age group in United States from February 12 to March 16. \label{fig:age_ICU}]{
     \includegraphics[scale=0.55]{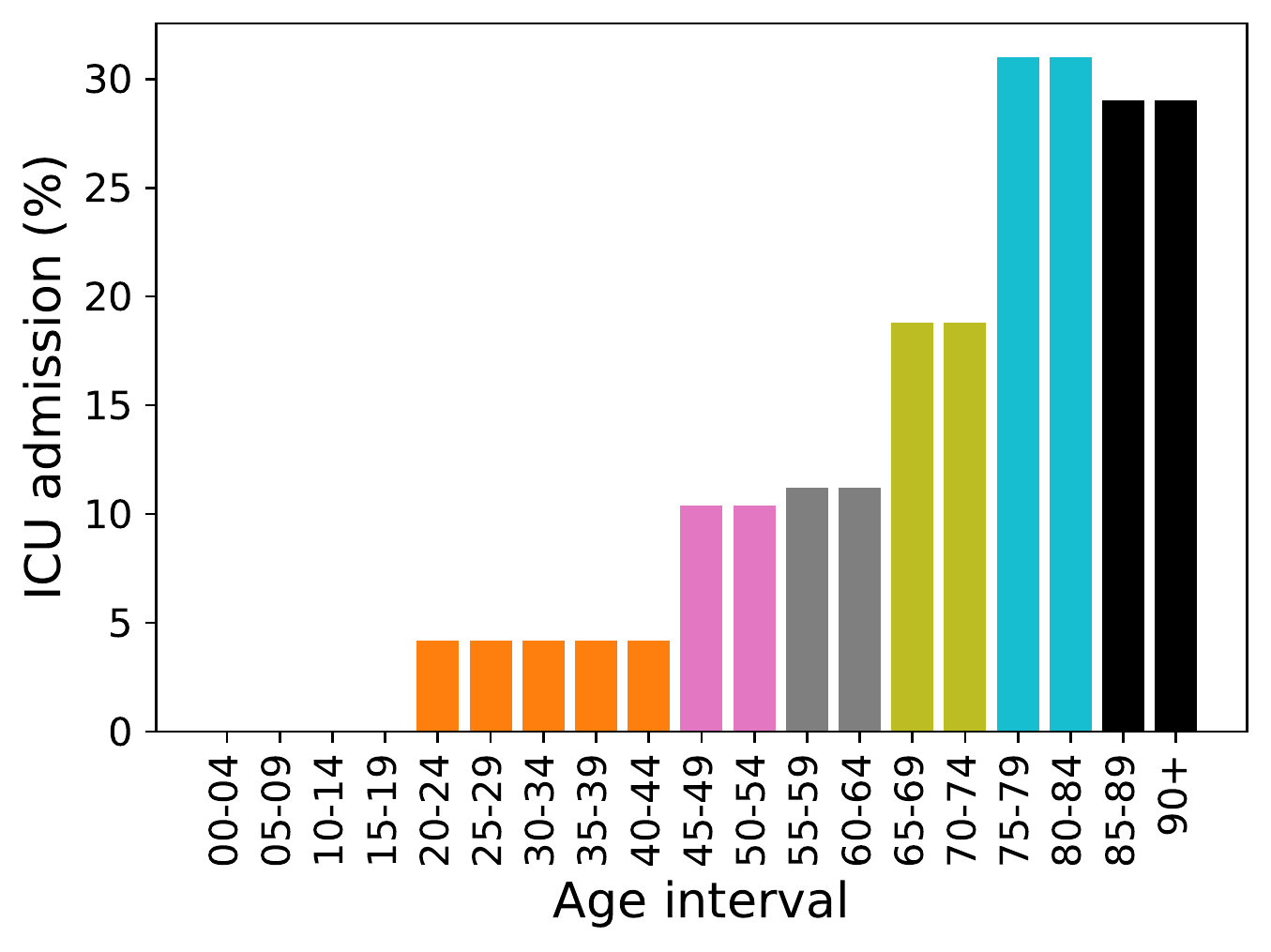}}

\caption{Data (until May 17, 2020) and model projections for both scenarios from Table \ref{tab:scenarios} for the state of S\~{a}o Paulo and the city of S\~{a}o Paulo\cite{brasil_io}.  Age-pyramid and age-dependent ICU admission. }
\label{fig:FITs_state_city_SP}
\end{figure*}

We estimate $\alpha$ and $t_0$ by  linear regression of $\log (\hat{n}^{*} / {\hat{n}(t)} - 1  )$, discarding the first $45$ days since case one.  The state of S\~{a}o Paulo has a population of $N= 46,289,333$ \cite{IBGE_age_SP}, while the city has $N= 12,252,023 $ inhabitants \cite{IBGE_city_SP} . 
The expected epidemic development for the two scenarios can be viewed in Figure \ref{fig:FITs_state_city_SP}. 


Figure \ref{fig:FITs_city_SP} makes explicit the grave situation of the city of S\~{a}o Paulo. Subnotification of cases and notification delays are not taken into account. Following current trends, we expect that both scenarios will soon become obsolete. 

Studies report that COVID-19 agravation are age-dependent \cite{Ioannidis2020.04.05.20054361,Davies2020.03.24.20043018, Modi2020.04.15.20067074}, making the age-pyramid central to the task of estimating demand for ICU beds.  We suppose that confirmed cases, both for the state and for the city of S\~{a}o Paulo, follow the  age-pyramid \cite{IBGE_age_SP} as shown in Figure \ref{fig:age_demography}. 

After age-structured sampling of cases, we sample over age-dependent probabilities of ICU admission following. For that we use data by age group reported for United States from February 12 to March 16 \cite{CDC} (see Figure \ref{fig:age_ICU}). These data might be preliminary, however ICU admission probability for individuals under 60 is clearly non-negligible. 

To build our estimates we assume that a individual remains in ICU for $ \tau= 14$ days before being removed. Given an specific day $t_j$, the total ICU beds demand on this day is given by 

\begin{equation}
 \hat{N}^{\text{ICU}} (t_j) = \hat{n}^{\text{ICU}}_{t_j} +   \hat{n}^{\text{ICU}}_{t_{j} -1} +  \dots  +   \hat{n}^{\text{ICU}}_{t_{j} - \tau}  \;,
\end{equation}
where $  \hat{n}^{\text{ICU}}_{t_j}$ is the Monte Carlo estimation obtained from sampling $\hat{n}(t_j)$, Eq.(\ref{eq:sig}), from age-structure probabilities, followed by sampling from ICU admission rates. 

We also introduce a multiplicative constant $S$, which accounts for subnotification of cases. This constant  is found by fitting public ICU occupation data to the median of the proposed scenarios and it is assumed to hold for all simulations throughout this work. Unfortunately, ICU bed occupation data is also not widely available in Brazil \cite{folha_freire}.

\section{Results for the State of S\~{a}o Paulo}

ICU occupation for the state of S\~{a}o Paulo is reported occasionally by the government on social networks \cite{twitter08, twitter09, twitter10, twitter11, twitter12, twitter13, twitter14, twitter15, twitter16, twitter17}. These scarce data points are represented by the blue circles in Figure \ref{fig:ICU_state_SP_DF1_DF2} together with both scenarios. Models are fitted to data up to May 17, 2020 and results until this date are depicted in gray.

Public data indicates that the state of S\~{a}o Paulo has 5934 COVID-19 ICU beds (availability on 05/22; calculated from \cite{twitter22}). We can thus find the time interval for system collapse, that is shown in Table \ref{tab:results_state_SP}. The table also shows demand peaks for each scenario and the subnotification factor $S$.

\begin{table*}[h!]
\small
\centering
\small
\centering
\begin{tabular}{| c c c c c c c |}
\hline
    & Collapse (68\% CI) & Collapse (95\% CI) & Max date & Max value (68\% CI) & Max value (95\% CI) &   $S$ \\
\hline\hline
Spain & 05/21 to 05/31 & 05/18 to 07/06 & 06/07 & 7154 $\pm$ 987 & 7154 $\pm$ 1934 & - \\
\hline
Median & 05/22 to 06/20 &  05/18 to 06/28 & 06/05 & 6180 $\pm$ 937 & 6180 $\pm$ 1836   &  2.1\\
\hline
Switzerland & 05/25 to 06/08  &  05/19 to 06/17 & 06/01 & 5389 $\pm$ 861 & 5389 $\pm$ 1688  & -\\
\hline
\end{tabular}
\caption{ \label{tab:results_state_SP} Relevant model values predictions: state of S\~{a}o Paulo. Collapse intervals assuming 5934 ICU beds available (on 05/22; from \cite{twitter22}). "Max date": date when average values reach the peak. "Max value": value for these dates.}
\end{table*}


 Curves like the ones in Figure \ref{fig:ICU_state_SP_DF1_DF2} could be constructed considering only the demand in the public health system (SUS), since the SUS dependent population is known for each and every Brazilian state \cite{UFMG_tec}. However, we lack  reliable data to estimate the subnotification factor $S$ and  the number of SUS ICU beds available for COVID-19 is not clear.

At the time of writing, we were able to extract ICU bed occupation from social networks for eight days after May 17 \cite{twitter19, twitter20, twitter21, twitter22, twitter23, twitter24, twitter25, twitter26}. In Table \ref{tab:predictions_state_SP} we use these data points to verify the quality of our predictions. We observe that data points are mostly compatible with the intervals suggested by the scenarios. 

\begin{table}
\small
  \centering
  \begin{tabular}{l*{17}{c}}
    \toprule
    & \multicolumn{3}{c}{68\% CI}  &  \multicolumn{1}{c}{} \\
    \cmidrule(lr){2-4}
    & \multicolumn{1}{c}{Spain} & \multicolumn{1}{c}{Median}  & \multicolumn{1}{c}{Switzerland}
    & \multicolumn{1}{c}{Data}\\
    \midrule
    
    05/19    & 4976 & 4801 & 4627  & 3659        \\
    
        &  $\pm$ 542 &  $\pm$ 676 &  $\pm$ 672 &   &   \\
    \addlinespace
    
   05/20    & 5142 & 4946 & 4750  & 4169        \\
    
        &  $\pm$ 697 &  $\pm$ 695 &   $\pm$ 694   &   &   \\
    \addlinespace

   05/21    & 5531 & 5083 & 4846 & 4224        \\
    
        &  $\pm$ 709 &  $\pm$ 705 &  $\pm$ 700  &   &   \\
    \addlinespace

   05/22    & 5489 & 5212 & 4934   & 4433        \\
    
        &  $\pm$ 726 &  $\pm$ 723 &  $\pm$ 719  &   &   \\
    \addlinespace        

       05/23    & 5653  & 5346  & 5039   & 4674       \\
    
        &  $\pm$ 746 &  $\pm$ 739 &  $\pm$ 731  &   &   \\
            \addlinespace
05/24    & 5811  & 5468  & 5125   & 4661       \\
    
        &  $\pm$ 759 &  $\pm$ 754 &  $\pm$ 749 &   &   \\
         \addlinespace
        
        05/25    & 5976  & 5582  & 5189   & 4283       \\
    
        &  $\pm$ 775 &  $\pm$ 767 &  $\pm$ 760 &   &   \\
        
     \addlinespace
        
        05/26    & 6124  & 5688  & 5251   & 4779       \\
    
        &  $\pm$ 792 &  $\pm$ 786 &  $\pm$ 780 &   &   \\

    \bottomrule
  \end{tabular}
  \caption{ State of São Paulo: Comparing model and  data \cite{twitter19, twitter20, twitter21, twitter22, twitter23, twitter24, twitter25, twitter26}. }
  \label{tab:predictions_state_SP}
\end{table}

\section{Results for the City of S\~{a}o Paulo}

Our sampling estimation method is quite simple and can be applied to any population once ICU occupation data is available.

\begin{figure*}[!htbp]
\subfigure[State of S\~{a}o Paulo. Shaded areas denote Monte Carlo errorbars. Detailed figures are depicted in table \ref{tab:results_state_SP}. \label{fig:ICU_state_SP_DF1_DF2}]{
\includegraphics[scale=0.55]{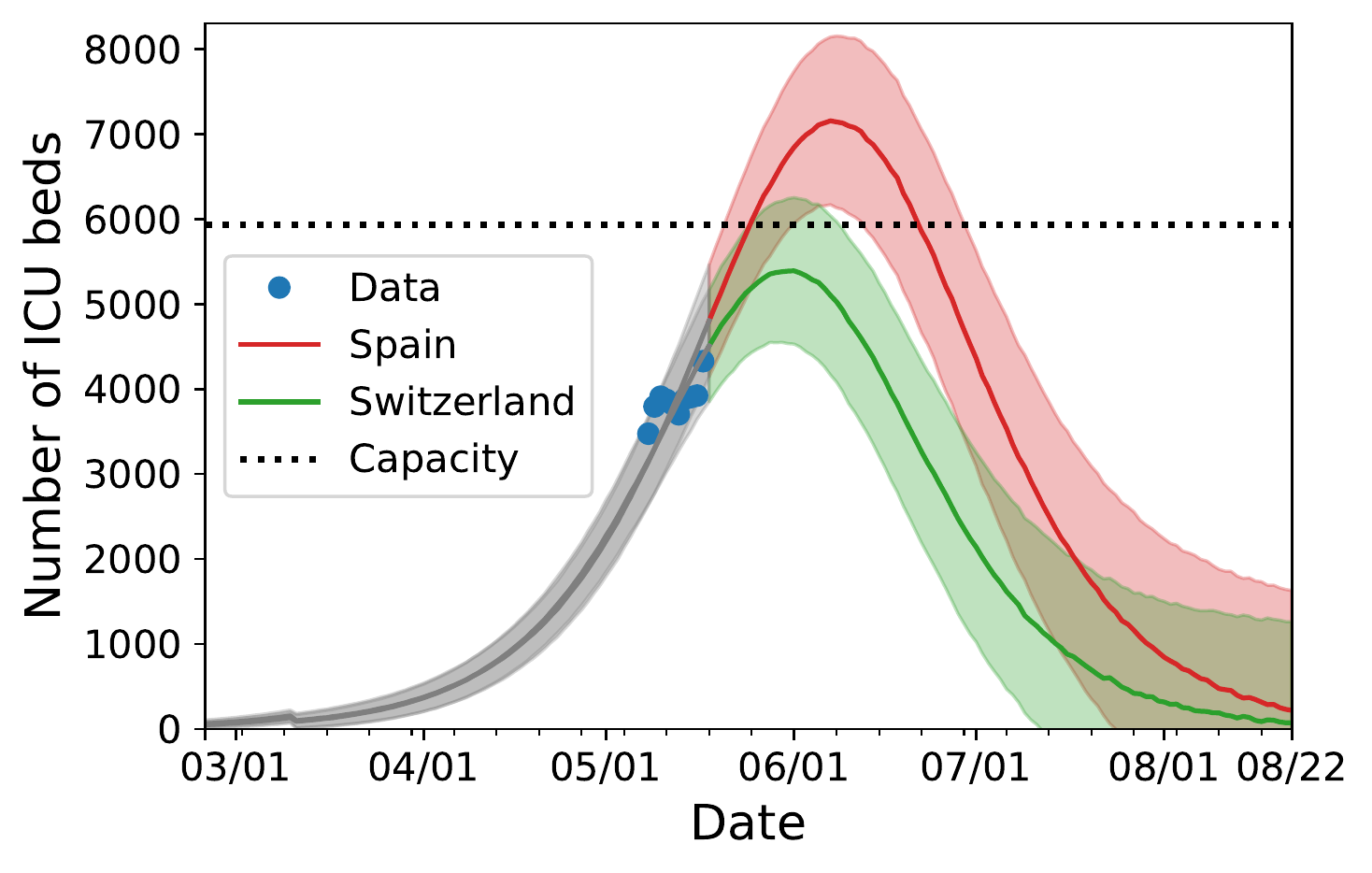}}
\subfigure[City of S\~{a}o Paulo. Shaded areas denote Monte Carlo errorbars. Detailed figures are depicted in table \ref{tab:results_city_SP}. \label{fig:ICU_city_SP_DF1_DF2}]{
 \includegraphics[scale=0.55]{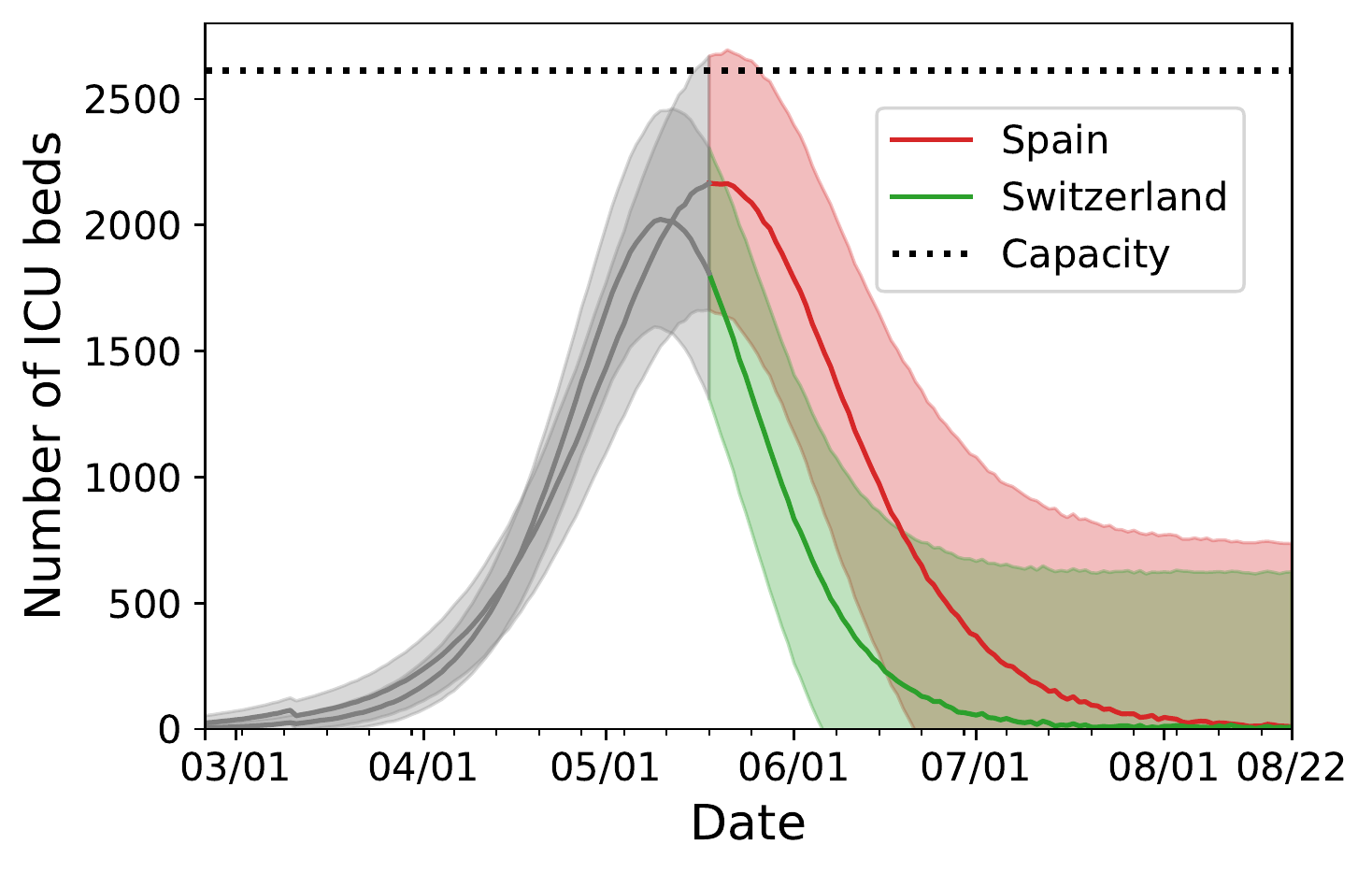}
    }    
    \label{fig:ICU_SP_state_}
    \caption{Scenarios for the state and city of S\~{a}o Paulo.}
\end{figure*}

The city of S\~{a}o Paulo is country's pandemic epicenter.  Validation will be possible only when reliable data is available. Unfortunately only occupation percentages are reported, but it is not clear how many ICU beds in total are available. Thus we are unable to find the subnotification factor $S$. Taking $S = 2.1$ from Table \ref{tab:results_state_SP} as a proxy,  we produced  Figure \ref{fig:ICU_city_SP_DF1_DF2}. ICU beds availability for COVID-19 is assumed to be 2611 \cite{boletim} on May 18,2020.


 We, unfortunately, cannot assure any reliability of these numbers, altogether with curves in Figure \ref{fig:ICU_city_SP_DF1_DF2} due to the lack of data for validation. We hope the necessary data will be eventually available. The dates when average values in each scenario reach the peak seem to indicate that the city of S\~{a}o Paulo would have already reached epidemic peak (see Table \ref{tab:results_city_SP}). However, data on daily new cases seems to indicate otherwise, pointing towards a situation worsening beyond the worst scenario employed.

\begin{table}[h!]
\small
\centering
\begin{tabular}{| c c c c |}
\hline
    &  Max date&  68\% CI &  95\% CI  \\
\hline\hline
Spain &  05/18 & 2166 $\pm$ 504 & 2166 $\pm$ 988 \\
\hline
Median &  05/15 & 2033 $\pm$ 474 & 2033 $\pm$ 929 \\
\hline
Switzerland & 05/10 & 2023 $\pm$ 431 & 2023 $\pm$ 845 \\
\hline
\end{tabular}
\caption{ \label{tab:results_city_SP} Relevant model values predictions: city of S\~{a}o Paulo.}
\end{table}

\section{Concluding remarks }

The number of confirmed cases is modelled by a logistic function, describing an initial exponential increase followed by a plateau. The exponential increase rate is estimated from available data.  Epidemic plateau is estimated from scenarios based on the dynamics observed on other countries. We employ limited ICU occupation data to estimate a subnotification factor and then use age-structured estimates to project scenarios for the progression of ICU demand.

 Information is  critical to deal with a sanitary crisis as we are facing. It should be government responsibility to use its resources to protect taxpaying citizens. Fortunately, we still have science as a guide.

\section{Acknowledgement}
R. Veiga is financially supported by CNPq under process 162857/2017. This work is also supported by the Covid Radar initiative.

\bibliographystyle{eplbib}	

\bibliography{main}		

\end{document}